\documentclass[showpacs,prl,twocolumn,groupedaddress,amsmath]{revtex4}
\usepackage{epsfig}
\usepackage{bm}

\begin{document}
\title{Magnetically Hidden Order of Kramers Doublets in $d^1$ Systems:
Sr$_2$VO$_4$}
\author{George Jackeli}
\altaffiliation[]{Also at Andronikashvili Institute of Physics, 0177 
Tbilisi, Georgia}
\author{Giniyat Khaliullin}
\affiliation{Max-Planck-Institut f\"ur Festk\"orperforschung, 
Heisenbergstrasse 1, D-70569 Stuttgart, Germany}
\begin{abstract} 
We formulate and study an effective Hamiltonian for 
low-energy Kramers doublets of $d^1$-ions on a square lattice.
We find that the system exhibits a magnetically hidden order in which 
the expectation values of the local spin and orbital moments both vanish.
The order parameter responsible for a time-reversal symmetry breaking has 
a composite nature and is a spin-orbital analog of a magnetic octupole.
We argue that such a hidden order  is realized in the layered perovskite
Sr$_2$VO$_4$. 
\end{abstract}
\date{\today}
\maketitle 

The ions with odd number of electrons have ground states which are at least
doubly degenerate. The latter, known as Kramers degeneracy, is
protected by time-reversal symmetry \cite{Abr70}. In Mott insulators the 
exchange interactions between Kramers doublets often induce a time-reversal 
symmetry breaking phase lifting the local degeneracy. As a result, 
a magnetic state with a well defined ordered pattern of local magnetic moments
is formed at low temperatures, which, in turn, leads to the new Bragg peaks 
in neutron scattering experiments.

The above conventional picture may, however, fail for the insulating systems
in which the local and non-local interactions between magnetically active
degrees of freedom compete with each other. This competition may lead 
to a spontaneous time-reversal symmetry breaking through the development 
of a more complex order parameter, while the magnetic dipole moments of 
spin and orbital origin, being locally entangled, would remain disordered 
across the transition. The resulting order can be thus hidden
to some experimental probes.

In this Letter we  show that such a magnetically hidden order 
may be realized in transition-metal compounds. In the following we focus
on a system of $d^1$ ions with partially filled
$t_{2g}$ levels on a square lattice, like the layered insulating 
compound Sr$_2$VO$_4$ \cite{Cyr90,Mat05,Zho07}. Here, a square 
lattice of V$^{4+}$ ions is formed in the $ab$-plane by corner-shared 
VO$_6$ octahedra, elongated along the $c$-axis. Sr$_2$VO$_4$ undergoes 
a phase transition on cooling at around $T_{c}\simeq 100$~K.
The crystal structure remains tetragonal across the transition,
the $c/a$ ratio jumps to a somewhat higher value, and
the magnetic susceptibility  shows a sharp drop near $T_c$ \cite{Zho07}. 
It has been suggested that a (weakly) first-order transition to an 
antiferromagnetic and orbitally ordered state could be responsible 
for the observed anomalies  \cite{Zho07}. 
However, the  neutron scattering experiments were not
able to detect any magnetic order in the low temperature phase \cite{Cyr90}.
The nature of the ordered phase still remains experimentally unknown.
Theoretically, as a possible candidate, a nontrivial orbital-stripe order 
coexisting with collinear antiferromagnetic spin order with large unit 
cell has been proposed \cite{Ima05}.   
 
Here, based on the microscopic theory, we propose that an unusual 
symmetry-breaking phase, induced by spin-orbit coupling and compatible 
with tetragonal crystal symmetry, is realized in Sr$_2$VO$_4$. 
We argue that the staggered ordering of composite spin-orbital objects, 
{\it magnetic octupoles}, is responsible for the phase transition. 
There is no static order of either spin or orbital magnetic moments 
in the ground state, hence the absence of magnetic Bragg peaks. 
 
The octupolar order has thus far been discussed primarily in the 
context of $f$-electron systems \cite{Kur08,Mom06}. The peculiar physics 
of $t_{2g}$ orbitals in tetragonal compounds uncovered here 
shows that this unusual state is well hosted by 
$d$-electron systems, too.

\begin{figure}
\epsfysize=30mm
\centerline{\epsffile{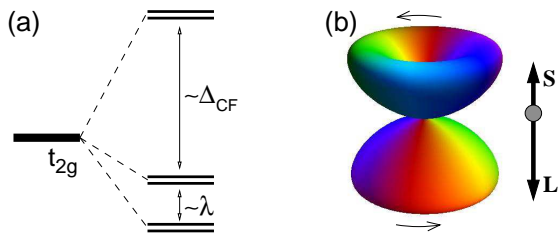}}
\caption{(Color online) (a) A schematic view of the local level structure:
the $t_{2g}$ level is split into three sets of Kramers doublets by
a tetragonal crystal field $\Delta_{\rm cf}$ and a spin-orbit
coupling $\lambda$. (b) Density profile of an electron in the ground
state doublet. The chirality of the  isospin up wave function, 
associated with the gradient of its phase, is shown by small arrows. 
The electron spin ${\vec S}$ and orbital angular momentum ${\vec L}$ are 
coupled antiparallel to each other.}
\label{fig1}
\end{figure}

{\it Low-energy effective Hamiltonian}.--  We first introduce the local degrees
of freedom and then discuss the exchange interactions between them. The 
V$^{4+}$ ion has a single unpaired electron residing in the $t_{2g}$ manifold of 
$xy$, $xz$, and $yz$ orbitals. The tetragonal elongation of the  oxygen 
octahedra  along the $z\!\parallel\!c$-axis only partly  lifts the threefold 
orbital degeneracy: The $xy$ orbital is pushed to a higher energy, while $xz$ 
and $yz$ orbitals remain degenerate. The orbital angular momentum is 
unquenched and the spin-orbit coupling is active. We thus start with a 
local Hamiltonian 
$H_{0}=\Delta_{\rm cf} (\frac{2}{3}-l_z^2)-\lambda {\vec l}\cdot{\vec S}$ 
consisting of a tetragonal crystal field $\Delta_{\rm cf}$ and a spin-orbit 
coupling $\lambda$. Here ${\vec S}$ is an electron spin, and $l=1$ is an 
effective angular momentum with  $|l_z\!\!=\!\!0\rangle \equiv \!\!|xy\rangle$, 
$|l_z\!\!=\!\!\pm 1\rangle \equiv 
\!\!-\frac{1}{\sqrt{2}}(i|xz\rangle\pm|yz\rangle)$. 
The total magnetic moment ${\vec M}=2{\vec S}+{\vec L}$, where 
$\vec{L}=-\kappa {\vec l}$ is a true angular momentum and $\kappa$  is a 
so-called covalency factor of order one \cite{Abr70}. The eigenstates 
of $H_{0}$ are spanned by three sets of Kramers doublets. The corresponding 
local level structure is schematically shown in Fig.~\ref{fig1}(a).  
Concerning the energy scales involved, the {\it ab initio} study 
of Sr$_2$VO$_4$ electronic structure suggests $\Delta_{\rm cf}\simeq 80$~meV 
\cite{Ima05}, and $\lambda\simeq 30$~meV for free V$^{4+}$ ion
is known experimentally \cite{Abr70}. In what follows, we retain only the
two low-energy levels and neglect the one located at high energy 
$\sim\Delta_{\rm cf}$. We label Kramers partners within a doublet 
by isospin index $\tilde{\uparrow}$ and $\tilde{\downarrow}$ 
($s^{z}=\pm\frac{1}{2}$), while the two doublets are 
denoted by pseudoorbital index $\pm$ ($\sigma^{z}=\pm1$). 
The wave functions of the ground state doublet, $\sigma^{z}=+1$,  are 
\begin{equation}
|\tilde{\uparrow}\rangle_{+}=|+1,\uparrow\rangle~,~~
|\tilde{\downarrow}\rangle_{+}=|-1,\downarrow\rangle~.
\label{eq1}
\end{equation} 
The corresponding electron density profile has an axial shape, and the two
Kramers partners, apart from the direction of an electron spin,
are distinguished by a chirality associated with the gradient of their phases 
determined by orbital angular momentum [see Fig.~\ref{fig1}(b)].

The first excited level, located at an energy 
$\delta=\lambda+\frac{1}{4}(2\Delta_{\rm cf}-\lambda)(1/\cos2\theta+1)$, 
has the following eigenstates: 
\begin{eqnarray}
|\tilde{\uparrow}\rangle_{-}&=&\sin\theta |-1,\uparrow\rangle+\cos\theta
|0,\downarrow\rangle~,\nonumber\\ 
|\tilde{\downarrow}\rangle_{-}&=&\sin\theta |+1,\downarrow\rangle+
\cos\theta |0,\uparrow\rangle~,
\label{eq2}
\end{eqnarray} 
where $\tan2\theta=2\sqrt{2}\lambda/(\lambda-2\Delta_{\rm cf})$. 
In the limit $\Delta_{\rm cf}\gg\lambda$ of interest here, 
$2\theta\sim \pi$ and $\delta\sim\lambda$.

We now discuss the interactions between neighboring ions. The effective  
Hamiltonian is obtained by projecting the corresponding $t_{2g}$ 
superexchange model of Ref.~\cite{Kha05} onto the reduced Hilbert space 
spanned by the two low energy Kramers doublets (\ref{eq1}) and (\ref{eq2}). 
We first consider the dominant part of the Hamiltonian and discuss later the 
effects of a finite Hund's coupling. We find 
\begin{equation}
{\cal H}=J\sum_{\langle ij\rangle} ({\vec s}_i\cdot {\vec s}_{j}+\frac{1}{4})
(1\pm\sigma^{x}_{i})(1\pm\sigma^{x}_{j})-\frac{\delta}{2}\sum_{i}\sigma^{z}_i~,
\label{eq3}
\end{equation}
where the first term describes the exchange coupling between the 
neighboring states, the $SU(2)$ isospin degrees of freedom 
are represented by $\vec s_i$, and Pauli matrices 
$\vec \sigma_i$ denote pseudo-orbitals, referred to as simply orbitals from 
now on (not to be confused with original $t_{2g}$ orbitals).
The upper (lower) sign is taken for a bond along $a(b)$ 
direction, $J=t^2/U$, where $t$ is a transfer integral and $U$ 
stands for the Coulomb repulsion. The local level splitting $\delta$ between 
two Kramers doublets is given by the second term. 

{\it The ground state properties}.-- The interaction between the isospins 
$\vec s_i$ in Eq.~(\ref{eq3}) depends on the orbital occupations through 
the operator $(1\pm\sigma^x_i)(1\pm\sigma^x_j)$. Since this operator is 
non-negative, the isospin exchange is purely antiferromagnetic (or equal 
to zero). This suggests that in the classical limit, 
$\langle {\vec s}_i{\vec s}_j\rangle=-\frac{1}{4}$, the expectation
value of the first term in the Hamiltonian vanishes and all orbital 
configurations are degenerate. This extensive classical degeneracy, 
inherent to the coupled spin-orbital systems \cite{Kha05}, is lifted here 
by the second term of the Hamiltonian \cite{Jac07} selecting  
uniform orbital order with $\sigma^{z}=+1$. The antiferromagnetic 
ordering of isospins $\vec s_i$ is then stabilized.

While the ground state in terms of staggered order of isospins may, 
at a first glance, look conventional, its physical properties are 
in fact very unusual and depend on the spatial orientation of the 
order parameter 
${\vec m}\equiv\langle2{\vec s}_{\bf Q}\rangle$ [${\bf Q}=(\pi,\pi)$ is the
ordering wave vector]. To illustrate this, we express the isospin 
${\vec s}$ in terms of the original, physical spin ${\vec S}$ and 
angular momentum ${\vec l}$, operators: 
\begin{equation}
s^{x(y)}=S^{x(y)}[(l^x)^2-(l^y)^2]~,~~ s^z=S^z~.   
 \label{eq4}
\end{equation}
The in-plane components of isospins are represented by a composite object, 
which is, remarkably enough,  a spin-orbital analog of a magnetic octupole. 
On the other hand, the axial component is equivalent to the physical spin. 
Thus, in-plane ordering of isospins (${\vec m}\perp z$) corresponds 
to a staggered order of magnetic octupoles, while the axial one 
(${\vec m}\parallel z$) corresponds to a magnetic dipolar order. 
In the former case, the local expectation values of both spin  and 
angular moments are exactly zero: 
$\langle{\vec S}_i\rangle=\langle{\vec l}_i\rangle=0$. 
This can be explicitly verified using the wave functions (\ref{eq1}) of 
the ground state doublet: the in-plane $g$-factor of this doublet 
$g_{ab}\equiv 0$. In the case of a dipolar order, ${\vec m}\parallel z$,  
we find again a somewhat unusual picture: the value of ordered magnetic 
moments is strongly suppressed, $\langle M^z\rangle=1-\kappa\ll 1$, 
because of compensation between spin and orbital magnetic moments. 
In the absence of covalency, i.e. $\kappa=1$, both $\langle M\rangle$ and 
$g_{c}=2(1-\kappa)$ vanish. 
       
The effective Hamiltonian (\ref{eq3}), with isospin rotational symmetry,
cannot select a direction of the order parameter ${\vec m}$: The latter 
can be  rotated from a purely dipolar character to a purely octupolar one
at no energy cost. In other words, a Goldstone mode describing the
out-of-plane rotation of isospins corresponds physically to fluctuations 
between dipolar and octupolar orderings. However, the $SU(2)$ isospin 
symmetry is only approximate and is broken by Hund's coupling $J_H$, neglected 
so far. Finite  $J_H$ induces the anisotropy term, which in the present case 
is of easy-plane form ${\cal H}_{\alpha}(i,j)=-\alpha J s_i^zs_{j}^{z}$,
with $\alpha=2J_{H}/U\ll 1$. It confines the isospins in the plane and selects
the staggered octupolar order, with vanishing dipolar moments 
$\langle{\vec S}_i\rangle$ and $\langle{\vec L}_i\rangle$ on every site. 
The emergence of this highly unusual state in an apparently simple 
$d^1$ Mott insulator like Sr$_2$VO$_4$ is surprising, given that its 
single-hole counterpart, $d^9$ perovskite La$_2$CuO$_4$, is a conventional 
antiferromagnet. The origin of new physics here is due to an unquenched 
spin-orbit coupling, the significance of which is being recognized in the 
context of various phenomena \cite{Tch04,Gan08,Jac09,Che09,Shi09}.

{\it Excitation spectra}.-- We now turn to the excitation spectrum above the
octupolar ordered state. We first study the 
Hamiltonian (\ref{eq3}) 
at a mean-field level, decouple it into isospin and orbital sectors, and 
discuss later the interactions between them. We employ isospin (orbital) 
wave representation for ${\vec s_i}$ (${\vec \sigma_i}$) operators in terms 
of Holstein-Primakoff bosons $b_i$ ($a_i$) and diagonalize the harmonic 
part of the mean-field Hamiltonian.

\begin{figure}
\epsfysize=48mm
\centerline{\epsffile{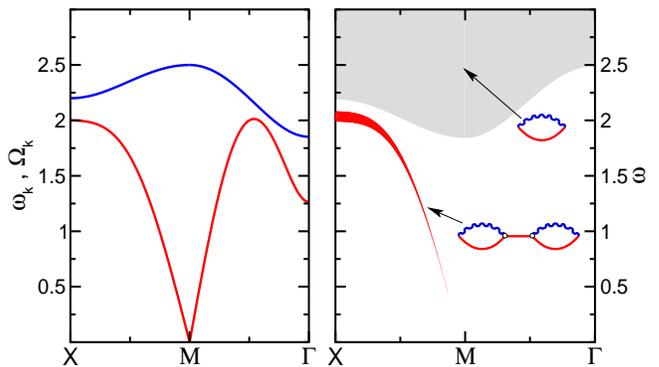}}
\caption{(Color online)  Left: The dispersions of elementary excitations 
along the direction $X(\pi,0)\rightarrow M(\pi,\pi)\rightarrow\Gamma$ in the
Brillouin zone. The lower (upper) curve corresponds to the isospin (orbital)
waves. Right: The spectrum of in-plane, $M^x$ component, magnetic 
excitations. It consists of a continuum of composite, isospin-orbital  
excitations (shaded area), and a quasiparticle part (lower curve, 
the width scales with intensity). The corresponding diagrams are also 
shown: The solid (wavy) lines denote the isospin (orbital) excitations 
and the open dots stand for the three-particle vertex in Eq.~(\ref{eq6}). 
Energies are given in units of $J$.} 
\label{fig2}
\end{figure}

The isospin (intradoublet) and orbital (interdoublet) excitation energies 
are given by $\omega_{\bf  k}=2J_s\sqrt{(1-\gamma_{\bf k}+\alpha\gamma_{\bf 
k})(1+\gamma_{\bf k})}$ and $\Omega_{\bf k}=\sqrt{\delta(\delta+8J_{o}
\gamma_{\bf k})}$, respectively.   
Here $\gamma_{\bf k}=\frac{1}{2}(\cos k_x+\cos k_y)$, 
$J_s=J\langle1+\sigma^{x}_{i}\sigma^{x}_{j}\rangle$, and  $J_o=J\langle {\vec 
s}_i{\vec s}_j+\frac{1}{4}\rangle$. At $T=0$, we estimate 
$J_s\simeq J$ and $J_o\simeq-0.08J$. The dispersion relations 
are plotted in Fig.~\ref{fig2} (left) for the realistic values of the
parameters ($\delta=2.2$ and $\alpha=0.2$ in units of $J$). The in-plane
isospin excitations are gapless, while the out-of-plane gap 
($\sim\sqrt{\alpha}$)  at $\Gamma$ point is induced by easy-plane 
anisotropy protecting octupolar order. The weakly dispersive 
interdoublet excitations are centered around $\delta$.
     
The above elementary excitations correspond, in fact, to the
fluctuations of rather unconventional degrees of freedom.
The in-plane isospin waves represent octupolar excitations, not
directly coupled to the conventional spectroscopic probes such as neutrons. 
To make some  predictions for  inelastic neutron scattering 
experiments, we now discuss the magnetic excitations. To this end, we express 
the local magnetic moment ${\vec M}_{i}$ in terms of isospin and orbital 
wave operators ($b_i$ and $a_i$, respectively):
\begin{eqnarray}
M^{x}_{i}&\simeq&(b_i+b_i^\dagger)(a_i+a_i^\dagger)~,
\;\;M^{y}_{i} \simeq e^{\imath{\bf Q}{\bf R}_i}(a_i+a_i^\dagger)~,\nonumber\\
M^{z}_{i}&\simeq&\imath(1-\kappa)e^{\imath{\bf Q}{\bf R}_i}(b_i-b_i^\dagger)~.
\label{eq5}
\end{eqnarray}
Here, only the leading order terms in boson operators are kept, and the
isospin order parameter along $y$-axis is assumed. All three 
components of ${\vec M}_{i}$ consist of fluctuating parts only. Hence, no 
magnetic Bragg peaks will be seen in elastic scattering. 
The inelastic response of the ordered state is rather nontrivial. 
The fluctuations of $M^{y}$ component,  which  have the largest spectral 
weight of order one, are given by orbital excitations only. We thus predict  
well-defined interdoublet excitations at energies $\sim\delta$ 
[upper curve in Fig.~\ref{fig2} (left)] to be observed by polarized 
neutrons. The out-of-plane magnetic response, $M^{z}$ component in 
Eq.~(\ref{eq5}), corresponds to  gapped out-of-plane isospin excitations.  
However, the  small spectral weight $\sim(1-\kappa)^2$
would probably make it hard to detect them. The in-plane $M^{x}$ response 
is given by the composite isospin-orbital excitations
[see Eq.~(\ref{eq5})]. The conventional spin wave 
excitations, inherent to magnetically ordered systems, are replaced 
here by a two-particle continuum [the shaded area in Fig.~\ref{fig2} 
(right)]. The latter is bounded from below by $\Omega_{{\bf k}+{\bf Q}}$, the
interdoublet excitation energy at momentum shifted by ${\bf Q}$.

The interaction between isospin and orbital waves, neglected so far,  
dynamically induce a linear coupling between in-plane magnetic and isospin
excitations. The three particle processes, responsible for this dynamic
hybridization, have the following form (in units of $J$):
\begin{equation}
{\cal H}^{\prime}=\sum_{{\bf k},{\bf q}}\{\eta_{\bf p}b^\dagger_{\bf 
q}b_{\bf k}+\eta_{\bf k}(b^\dagger_{-{\bf k}}b^\dagger_{{\bf q}}+b_{{\bf
k}}b_{-{\bf q}})\}(a_{\bf p}+a_{-{\bf p}}^\dagger )~,
\label{eq6}
\end{equation}
where the matrix element $\eta_{\bf p}=(\cos p_x-\cos p_y)$, and 
${\bf p}={\bf q}-{\bf k}$. The  second order perturbative correction from 
${\cal H}^{\prime}$ to the in-plane magnetic susceptibility $\chi_{xx}$ is
given by the lower diagram in Fig.~\ref{fig2} (right). When this correction is 
included, the spectral function of  in-plane magnetic excitations 
$\chi_{xx}^{\prime\prime}$ receives also a quasiparticle contribution. It
lies below the continuum and is given by 
\begin{equation}
\chi_{xx}^{\prime\prime}({\bf k},\omega)=\left(\frac{J\eta_{\bf k}}
{\delta+2J}\right)^2\sqrt{\frac{1-\gamma_{\bf k}}
{1+\gamma_{\bf k}}}\;\;\delta(\omega-\omega_{\bf k})~.
\label{eq7}
\end{equation}
Its intensity is peaked at $X$-point and dies out when approaching to
the ordering wave vector [see Fig.~\ref{fig2} (right)]. The Goldstone mode 
is thus invisible, reflecting the absence of magnetic Bragg peaks. 
Along $(1,1)$ direction, the hybridization matrix element 
vanishes and the quasiparticle peak is absent.  

\begin{figure}
\epsfysize=53mm
\centerline{\epsffile{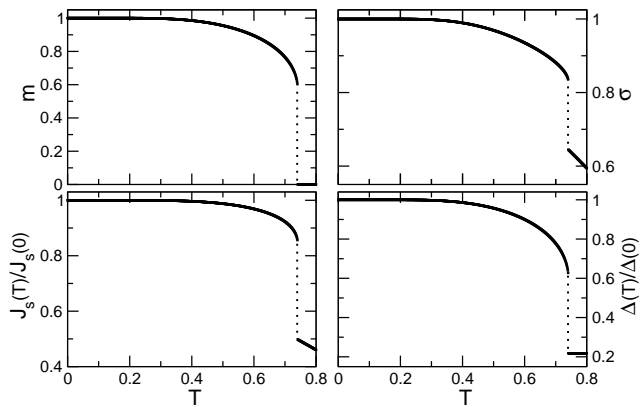}}
\caption{The isospin $m$ and orbital $\sigma$ order parameters, 
isospin exchange $J_s$, and orbital gap $\Delta$ vs temperature. 
The broken lines mark corresponding jumps at the first order transition.}
\label{fig3}
\end{figure}

{\it Finite temperature transition}.-- We now turn to finite temperatures 
and show that interplay between interdoublet and intradoublet excitations leads
to a first order phase transition. The in-plane ordering of isospins  breaks 
their rotational symmetry around $z$-axis. 
Note that  the  uniform order of low-energy doublets, 
$\sigma\equiv\langle\sigma^z_{i}\rangle=+1$, conserves the tetragonal crystal
symmetry, as  $xz$ and $yz$ orbitals are equally occupied on every site. 
The scale of the transition temperature is set by the isospin exchange 
energy $J_{s}$, which depends on orbital correlations. On the other hand, 
the orbital correlations themselves are determined by  isospin ones  
controlling the orbital exchange $J_{o}$. The nonlinear feedback effects 
between two degrees of freedom  can drive  a first order transition.

To substantiate the above picture, we employ a mean-field approach. The isospin
order parameter $m$ is given by a self-consistent equation $m=\tanh (mJ_s/T)$ 
with $J_s=J\langle1+\sigma^{x}_{i}\sigma^{x}_{j}\rangle$, where  
the orbital correlation function 
$\langle\sigma^{x}_{i}\sigma^{x}_{j}\rangle=\sum_{\bf k}\gamma_{\bf k}
\delta/[\Omega_{\bf k}\tanh(\Omega_{\bf k}/2T)]$. The energy of orbital
excitations $\Omega_{\bf k}$ is controlled by orbital exchange 
$J_{o}=J(1-m^2)/4$.  

Shown in Fig.~\ref{fig3} are the temperature dependences of isospin and 
orbital order parameters, together with isospin exchange and orbital 
excitation gap $\Delta$. The first order jumps are clearly seen in all 
physical quantities. This is compatible with the abrupt 
changes in lattice parameters and magnetic susceptibility observed 
experimentally near $T_c\simeq100$~K \cite{Zho07,note1}.  

Fig.~\ref{fig3} shows that not only isospin order parameter $m$
but also the exchange energy $J_s$ jumps at the transition. This suggests 
pronounced magnetoelastic effects and explains an abrupt in-plane contraction 
of the crystal, enhancement of a $c/a$ ratio. Note also that jumps in 
$\sigma$ and $\Delta$  imply an increase of the population 
of the low-energy doublet. As the latter does not include planar $xy$ 
orbital, Jahn-Teller coupling acts in the same direction  as magnetoelastic 
effect and further enhances a $c/a$ ratio. Finally, since the ground state 
doublet (\ref{eq1}) is non-magnetic, the magnetic susceptibility drops at the 
transition to the level determined by  the Van Vleck contribution from 
the transitions to the high energy doublets \cite{note2}. 

In conclusion, a magnetically hidden octupolar order may be induced 
by spin-orbit coupling in $d^1$ transition metal oxides. We have argued 
that  such a hidden order is realized in the perovskite Sr$_2$VO$_4$, 
explaining thereby the puzzling absence of magnetic Bragg peaks in this 
Mott insulator. The present theory suggests a nontrivial magnetic excitation 
spectrum in the ordered state, that can be verified by neutron scattering 
experiments. The spin-resolved circularly polarized photoemission experiment 
would be another test of the present scenario. This technique measures the 
${\vec l}\cdot{\vec S}$ scalar product \cite{Miz01}, which we predict to be
$\sim 0.5$ in the ground state. Finally, we suggest that another candidate 
to exhibit octupolar order is Sr$_2$NbO$_4$, in which a more pronounced 
spin-orbit coupling is expected. Sr$_2$NbO$_4$ is known to be a Mott 
insulator \cite{Isa01}, however, its low temperature magnetic properties 
have not yet been reported. 
 
We would like to thank B. Keimer, M. Haverkort, and P. Horsch 
for stimulating discussions. 



\begin{thebibliography}{20}
\bibitem{Abr70} A. Abragam and B. Bleaney, {\it Electron Paramagnetic 
Resonance of Transition Ions} (Clarendon Press, Oxford, 1970).

\bibitem{Cyr90} 
M. Cyrot {\it et al}., J. Solid State Chem. {\bf 85}, 321 (1990).

\bibitem{Mat05} J. Matsuno {\it et al}., 
Phys. Rev. Lett. {\bf 95}, 176404 (2005).

\bibitem{Zho07} H.D. Zhou {\it et al}., 
Phys. Rev. Lett. {\bf 99}, 136403 (2007).

\bibitem{Ima05} Y. Imai, I. Solovyev, and M. Imada, 
Phys. Rev. Lett. {\bf 95}, 176405 (2005).

\bibitem{Kur08} See, e.g., Y. Kuramoto, Prog. Theor. Phys. 
Suppl. {\bf 176}, 77 (2008), and references therein. 

\bibitem{Mom06} A few exceptions include the case of frustrated magnets 
[T. Momoi, P. Sindzingre, and N. Shannon, 
Phys. Rev. Lett. {\bf 97}, 257204 (2006); 
M.E. Zhitomirsky, Phys. Rev. B {\bf 78}, 094423 (2008)], and possible 
order of complex $e_g$ orbitals in ferromagnetic manganites 
[J. van den Brink and D. Khomskii, Phys. Rev. B {\bf 63}, 140416(R) (2001)].

\bibitem{Kha05} G. Khaliullin, 
Prog. Theor. Phys. Suppl. {\bf 160}, 155 (2005).

\bibitem{Jac07} At vanishing spin-orbit coupling, $\delta = 0$ and 
the quantum effects select the spin-singlet dimer states, see 
G. Jackeli and D.A. Ivanov, Phys. Rev. B {\bf 76}, 132407 (2007). 

\bibitem{Tch04} O. Tchernyshyov, Phys. Rev. Lett. {\bf 93}, 157206 (2004). 

\bibitem{Gan08} S. Gangadharaiah, J. Sun, and O.A. Starykh,  
Phys. Rev. Lett. {\bf 100}, 156402 (2008). 

\bibitem{Jac09} G. Jackeli and G. Khaliullin, 
Phys. Rev. Lett. {\bf 102}, 017205 (2009).

\bibitem{Che09} G. Chen, L. Balents, and A.P. Schnyder, 
Phys. Rev. Lett. {\bf 102}, 096406 (2009).

\bibitem{Shi09} A. Shitade {\it et al.}, Phys. Rev. Lett. 
{\bf 102}, 256403 (2009).

\bibitem{note1} Our mean-field study gives a  transition temperature 
$T_{MF}\sim 0.74 J$. Using $t=0.19$ eV and $U=2.58$ eV, 
suggested by {\it ab initio} calculations \cite{Ima05},  we estimate  
$J\simeq 14$ meV, and find a reasonable value $T_{MF}\sim 120$~K.

\bibitem{note2} The susceptibility upturn seen at low temperatures 
is possibly due to spin-one V$^{3+}$ impurities induced by the oxygen 
deficiency, see 
N. Suzuki, T. Noritake, and T. Hioki, Mater. Res. Bull. {\bf 27}, 1171 (1992).

\bibitem{Miz01} T. Mizokawa {\it et al}., 
Phys. Rev. Lett. {\bf 87}, 077202 (2001); G. Ghiringhelli {\it et al}., 
Phys. Rev. B {\bf 66}, 075101 (2002).

\bibitem{Isa01} 
K. Isawa and M. Nagano, Physica C {\bf 357}-{\bf 360}, 359 (2001).
\end{thebibliography}
\end{document}